\documentclass[12pt]{article}
\pdfoutput=1
\usepackage{putex}
\usepackage{amsmath}
\usepackage{amssymb}
\usepackage{graphicx}
\usepackage{epstopdf}
\usepackage{enumerate}
\usepackage{cite}
\usepackage{tensor}
\usepackage{slashed}
\usepackage{feynmf}
\usepackage{hyperref}
\usepackage{bbold}
\usepackage{dsfont}
\usepackage{mathrsfs}
\usepackage{caption}
\usepackage{subcaption}

\usepackage{youngtab}
\usepackage{amsfonts}
\usepackage{braket}

\newcommand\numberthis{\addtocounter{equation}{1}\tag{\theequation}}

\renewcommand{\theequation}{\thesection.\arabic{equation}}

\newcommand{\epst}{\tilde{\epsilon}}

\usepackage[utf8x]{inputenc}
\usepackage{titlesec}
\usepackage{fixltx2e} 
\usepackage{cleveref}
\hypersetup{breaklinks}

\usepackage[usenames,dvipsnames,svgnames,table]{xcolor}

\def\eqr{\eqref}

\newcommand {\be} {\begin {equation}}
\newcommand {\ee} {\end {equation}}
\newcommand{\bea}{\begin{eqnarray}}
\newcommand{\eea}{\end{eqnarray}}

\def\eps{\epsilon}

\def\rt{\rightarrow}

\newcommand{\fO}{O}
\newcommand{\tG}{\tilde{G}}
\newcommand{\D}{\Delta}
\newcommand{\lr}[1]{\left( #1 \right)}
\newcommand{\vex}[1]{\left\langle #1 \right\rangle}

\newcommand{\abs}[1]{\left| #1 \right|}

\def\eqr{\eqref}

 \newmuskip\pFqmuskip

\newcommand*\pFq[6][8]{%
  \begingroup 
  \pFqmuskip=#1mu\relax
  \mathcode`\,=\string"8000
  \begingroup\lccode`\~=`\,
  \lowercase{\endgroup\let~}\pFqcomma
  {}_{#2}F_{#3}{\left[\genfrac..{0pt}{}{#4}{#5};#6\right]}%
  \endgroup
}
\newcommand{\pFqcomma}{\mskip\pFqmuskip}

\makeatletter
\renewcommand{\@maketitle}{
\newpage
 \begin{center}%
  {\large\bfseries \@title \par}%
 \end{center}%
 \par} \makeatother

\numberwithin{equation}{section}

\titleformat*{\section}{\large\bfseries}

\begin{document}

\institution{UCLA}{ \quad\quad\quad\quad\quad\quad\quad\quad\quad Mani L. Bhaumik Institute for Theoretical Physics, \cr Department of Physics and Astronomy, University of California, Los Angeles, CA 90095, USA}

\title{Black holes from CFT:  \\ Universality of correlators at large $c$}

\authors{Per Kraus, Allic Sivaramakrishnan, and  River Snively}

\abstract{  Two-dimensional conformal field theories at large central charge and with a sufficiently sparse spectrum of light states have been shown to exhibit universal thermodynamics \cite{ Hartman:2014oaa}.  This thermodynamics matches that of AdS$_3$ gravity, with a Hawking-Page transition between thermal AdS and the BTZ black hole.  We extend these results to correlation functions of light operators.  Upon making some additional assumptions, such as large $c$ factorization of correlators, we establish that the thermal AdS and BTZ solutions emerge as the universal backgrounds for the computation of correlators.     In particular, Witten diagrams computed on these backgrounds yield the CFT correlators, order by order in a large $c$ expansion, with exponentially small corrections.  In pure CFT terms, our result is that thermal correlators of light operators are determined entirely by light spectrum data.   Our analysis is based on the constraints of modular invariance applied to the torus two-point function.       }

\date{}

\maketitle
\setcounter{tocdepth}{2}
\tableofcontents

\section{Introduction}

In classical physics, black holes have a clear meaning as representing the inevitable endpoint of the gravitational collapse of suitably dense configurations of matter.  In the quantum world, by contrast, the precise status of black hole solutions is a far more subtle question that continues to provoke debate; see e.g. \cite{Harlow:2014yka}.  A basic question is whether standard black hole geometries, realized in the quantum theory as coherent states of the gravitational field, are in some sense close to the actual solutions resulting from the collapse of matter in a particular microstate, or instead, do the actual solutions look dramatically different, with the usual black hole only arising as an effective description after some sort of coarse graining?  Some version of the latter scenario seems nearly inescapable if one demands that black hole evaporation be described by a unitary S-matrix governed by more-or-less ordinary laws of physics (no macroscopic violations of locality, etc.) \cite{Mathur:2009hf,Almheiri:2012rt}.

The AdS/CFT correspondence offers a framework to address such questions without resorting to ad hoc speculations.   In this paper we consider observables that are under good theoretical control, namely boundary correlation functions of low dimension operators computed in the thermal ensemble.  We pose the question: under what conditions are such correlation functions, at sufficiently high temperature, accurately computed by Witten diagrams in the standard black hole geometry?

We work in the context of the AdS$_3$/CFT$_2$ correspondence, which provides several technical advantages while retaining the essential physical elements present in higher dimensional examples.  Our guiding philosophy is that we wish to deduce the emergence of black hole physics while only making well-motivated assumptions about the CFT in the low energy and low temperature regime.  The key feature that allows us to proceed in this manner is modular invariance of the CFT partition function and correlators. A number of other works have used modular invariance to probe the AdS/CFT correspondence, for example \cite{Hellerman:2009bu,Belin:2017nze,Cardy:2017qhl,Maloney:2016kee,Kraus:2016nwo,Chang:2015qfa,Chang:2016ftb,Collier:2016cls,Collier:2017shs}. Our basic result can be stated as follows.  Under some mild assumptions corresponding to what one thinks of as a good holographic CFT at low energies, high temperature correlation functions computed at large central charge are indeed given, order by order in bulk perturbation theory, by Witten diagrams computed in the Euclidean BTZ geometry, with deviations being exponentially small in the central charge.\footnote{In this work we only consider one-point and two-point correlators, but we expect our results to extend to higher point correlators.}    To see a breakdown of the black hole geometry at the level of thermal correlators one therefore needs to either examine these exponentially small terms, or consider kinematical configurations, such as large Lorentzian time separations, that take one outside our assumptions.

Our work is the natural extension of the work of Hartman, Keller, and Stoica (HKS) \cite{ Hartman:2014oaa}.  HKS assumed a sufficiently sparse spectrum of states at energies below the black hole threshold, and then used modular invariance to deduce that at large $c$ the thermal free energy matches that of thermal AdS and BTZ in the low and high temperature regimes respectively.    We will assume the same sparseness conditions as HKS, and in addition make some assumptions about the strength of couplings in the low energy theory.  These assumptions are described in more detail in the next section.  We make reference to light (L), medium (M), and heavy (H) operators,\footnote{Note that our usage of light, medium, and heavy differs from that of HKS.} according to the value of  their conformal dimension $\Delta$.\footnote{$\Delta = h+\overline{h}$ is the total scaling dimension.}   Light operators obey $\Delta < \Delta_c$, where $\Delta_c $ is held fixed as $c\rt \infty$.   Medium operators lie in the range $\Delta_c < \Delta < c/12$, where the upper limit is the black hole threshold.  Heavy operators obey $\Delta > c/12$, and create black hole microstates.  Our approach is to compute a quantity defined with respect to a given $\Delta_c$, perform an asymptotic expansion in $1/c$, and then send $\Delta_c \rt \infty$. We make the usual holographic CFT assumption of large $c$ factorization, which is the statement that correlators of light operators admit an asymptotic expansion in powers of $1/\sqrt{c}$, and that the spectrum of such operators organizes into single-trace and multi-trace operators (a nice, general discussion may be found in \cite{ElShowk:2011ag}).  Our next assumption concerns the growth of light correlators computed in medium energy states; we assume that such correlators grow at most polynomially in the medium energy dimension, which, as we discuss, is the natural expectation.   This assumption is necessary to ensure that correlators of light operators computed at low temperature receive negligible contributions from states of energy $\Delta \sim c$, which would represent a breakdown of low energy effective field theory. Finally we make a technical assumption regarding the expansion of low temperature thermal correlators.

We then proceed to study the implications of modular invariance on one-point and two-point thermal correlators.   The one-point function is not so easy to study in isolation, because when expanded as a sum over states the terms in the sum have no definite sign, making it hard to deduce bounds.  So we instead focus first on the two-point function, which does admit an expansion in terms of positive quantities, and then later circle back to the one-point function using these results.  As already stated, our main result is that in the high temperature regime, $T > 1/2\pi$, these correlators, to all orders in the $1/\sqrt{c}$ expansion, are computed from Witten diagrams in the Euclidean BTZ geometry. Thermal one-point functions have a typical size of $1/\sqrt{c}$.  From the bulk point of view, this is the statement that scalar fields vanish in the classical limit, which we can think of as a version of a no-hair theorem derived from CFT considerations.  Of course, such a result is not surprising given that HKS already established that the thermodynamics of the CFT is in universal agreement with the hairless BTZ solution.  We should also emphasize that this version of the no-hair statement concerns solutions that dominate in the canonical ensemble; it allows for the existence of novel solutions that dominate at fixed energy rather than fixed temperature.

The rest of this paper is organized as follows. In section 2, we review the HKS argument that we extend to prove our main result. In section 3 we state and motivate the assumptions necessary for our argument. In section 4, we use the modular bootstrap to prove one of our main results: under our assumptions, thermal two-point functions of light operators are determined by the light sector of the theory to all orders in an asymptotic expansion in $1/\sqrt{c}$. In section 5, we show that our two-point function result implies that thermal one-point functions of light operators are also determined by the light sector in the same way. We conclude with a brief discussion of possible extensions of this work.    Appendix A contains certain calculational details  that we omitted in section 4.

\section{Review of the HKS argument}\label{HKSrev}

In this section we review results from ref. \cite{Hartman:2014oaa} for the partition function of a 1+1-dimensional large-$c$ CFT living on a circle. We first divide the spectrum into light (L) and heavy (H) states,
\begin{equation}
{L} = \left\{ E\leq {\epsilon}  \right\},
~~~~~~~~~
{H} = \left\{ E > {\epsilon}  \right\}.
\end{equation}
However, we bring to the reader's attention that in all other parts of this paper we use a different definition of ``light" (as in the introduction); in this section only we are adopting the terminology in HKS. Energy is related to scaling dimension as
\begin{equation}
E = \Delta - {c\over 12}.
\end{equation}

The $L, H$ contributions to the partition function are
\begin{equation}
Z_L = \sum_L e^{-\beta E}, ~~~~~~~   Z_H = \sum_H e^{-\beta E},
\end{equation}
where $\beta$ is the inverse temperature and the sum is over all states in the relevant range.
HKS use modular invariance to show that at large $c$ a theory with a sufficiently sparse light spectrum obeys  $\log Z \approx \log Z_L \approx \beta c/12$ at temperature $\beta >2\pi$.   The precise meaning of $\approx$ will become apparent.

 The $L, H$ contributions to the modular-transformed partition function ($\beta \rightarrow \beta' = 4\pi^2/\beta$) are denoted
\begin{equation}
Z_L' = \sum_L e^{-\beta' E}, ~~~~~~~   Z_H' = \sum_H e^{-\beta' E}.
\end{equation}
Modular invariance implies
\begin{equation}
Z_L-Z'_L = Z'_H-Z_H.
\label{HKS2.5}
\end{equation}
What $Z_L$ gains under modular transformation, $Z_H$ loses.

Assume  $\beta > 2\pi$. We want a bound on $Z_H= Z_H'-Z_L+Z_L'$ relative to $Z_L$. Because every term in the partition function sum is positive, $Z_H$ can be bounded in terms of $Z_H'$,
\begin{equation}
Z_H = \sum_H e^{(\beta'-\beta) E} e^{-\beta' E} \leq
e^{(\beta'-\beta)\epsilon} Z_H'.
\label{HKS2.6}
\end{equation}
This implies a bound on $Z_H-Z_H'$,
\begin{equation}
Z_H-Z_H' \leq Z_H' (e^{(\beta'-\beta)\epsilon}-1).
\label{HKS2.7}
\end{equation}
We can now use modular invariance \eqref{HKS2.5} to exchange $Z_H-Z_H'$ for $Z_L'-Z_L$ to obtain a bound on $Z_H'$ in terms of $Z_L$,
\begin{equation}
Z_H'
\leq
(1-e^{(\beta'-\beta)\epsilon})^{-1}(Z_L-Z_L')
\leq
(1-e^{(\beta'-\beta)\epsilon})^{-1}Z_L.
\numberthis
\label{HKS2.8}
\end{equation}
According to \eqref{HKS2.6} this bound translates into a bound on $Z_H$ in terms of $Z_L$.
\begin{equation}
Z_H \leq
\frac
{e^{(\beta'-\beta)\epsilon}}
{1-e^{(\beta'-\beta)\epsilon}} Z_L.
\label{HKS2.9}
\end{equation}
It follows that
\begin{equation}
\ln Z_L \leq \ln Z = \ln (Z_L + Z_H) \leq \ln Z_L - \ln(1-e^{(\beta'-\beta)\epsilon}).
\label{HKS2.10}
\end{equation}
Using modular invariance, one obtains a similar expression for $\beta < 2\pi$. So far, everything holds for an arbitrary compact, unitary, CFT.

We now consider a family of CFTs with a large $c$ limit.  Taking  $\epsilon$ to be independent of $c$, we then have that  at large $c$,
\begin{equation}
\log Z =
\left\{
\begin{matrix}
 \log Z_L + \mathcal{O}(c^0),\quad \beta >2\pi, \\
\log Z_L' + \mathcal{O}(c^0),\quad \beta < 2\pi~.
\end{matrix}  \right.
\label{HKS2.11}
\end{equation}
Next, it's easy to see that $\log Z_L = \beta c/12 + {\cal O}(c^0)$ provided that the following sparseness condition is obeyed,
\begin{equation}
\rho(E) \lesssim e^{2\pi(E+\frac{c}{12})} = e^{2\pi \Delta} ~~~~ E \leq \epsilon,
\label{HKS2.14}
\end{equation}
where $\lesssim$ is defined in footnote \ref{ftsparse}.  We will have more to say about this assumption in the next section.
Under this assumption we then have
\begin{equation}
\log Z =
\left\{
\begin{matrix}
 \beta c/12 + \mathcal{O}(c^0),\quad \beta >2\pi, \\
\beta' c/12 + \mathcal{O}(c^0),\quad \beta < 2\pi~.
\end{matrix}  \right.
\label{HKS2.12}
\end{equation}
This result for the partition function implies a Cardy density of states for $E>c/12$,
\begin{equation}
\rho(E) \approx e^{2\pi \sqrt{{c\over 3}E}}, ~~~~~ E>c/12,
\end{equation}
where the smooth function $\rho(E)$ is obtained by averaging the microscopic density of states over a small energy window.

$\rho(E)$ is  non-universal for $0<E<c/12$. However, the assumption \eqref{HKS2.14} implies  $\log \rho (E) \leq \pi c/6+2\pi E$ in this range, and at large $c$. These states never dominate in the canonical ensemble.

The main takeaway message is that the sparseness assumption together with modular invariance at large $c$ implies a universal result for the leading behavior (in $c$) of the canonical partition function, and this result furthermore matches the partition function obtained from AdS$_3$ gravity in the classical limit.

\section{Assumptions}\label{sumps}

In this section we lay out the precise assumptions that will be invoked in our correlation function analysis, and  discuss the motivation for these assumptions, which come from expectations about which properties we expect of a CFT with a ``good" holographic dual.  In stating our assumptions we are thinking in terms of a family or sequence of CFTs such that we can take $c \rt \infty$ within this space of CFTs.  Each member of the family is assumed to be a compact, unitary, CFT.

In the following, light, medium, and heavy operators are defined as having scaling dimensions:
\bea
L= \{0 \leq \Delta \leq \Delta_c\}~,\quad  M= \{\Delta_c < \Delta \leq  {c\over 12}+\eps\}~,\quad H=\{ \Delta > {c\over 12}+\eps\}~.
\eea
Here $\eps$ is held fixed as $c \rt \infty$.  The cutoff $\Delta_c$ is taken to infinity {\em{after}} performing the large-$c$ expansion.

\vspace{.2cm}
\noindent
{\bf{Assumption 1: Sparse spectrum}}
\vspace{.2cm}

This condition is widely discussed in the AdS/CFT literature, e.g. \cite{Heemskerk:2009pn,ElShowk:2011ag}, and a specific version of it was noted in the last section.
At the crudest level, we should demand that as $c \rt \infty$ the number of local operators with dimension below any fixed value should remain finite.  A diverging number of operators could invalidate the usual loop expansion in the bulk: a growing number of light fields running in loops could make loop diagrams compete with  or dominate over tree diagrams.  HKS assume a specific version of this statement, namely that the density of operators at dimension $\Delta$ should obey\footnote{Following HKS, the inequality $e^x \lesssim e^y$ means that $\lim x/y \leq 1$.  So, for example, the right hand side of the inequality in (\ref{hksineq}) could be multiplied by a polynomial in $\Delta$.\label{ftsparse} }
\be\label{hksineq}
\rho(\Delta)  \lesssim e^{2\pi \Delta}~,\quad  \Delta \leq {c\over 12} +\epst~.
\ee
The quantity $\epst$ is eventually taken to zero; we distinguish it from the quantity $\epsilon$ introduced below, which remains finite at large $c$.

As reviewed in the previous section, HKS showed that their sparseness assumption on the light spectrum implied that
\be\label{hksineq2}
\rho(\Delta)  \lesssim e^{2\pi \Delta}~,\quad  \Delta \leq {c\over 6}~,
\ee
as well as
\be\label{cardyspec}
\rho(\Delta)  \approx e^{2\pi \sqrt{{c\over 3}(\Delta-{c\over 12})}}~,\quad  \Delta > {c\over 6}~.
\ee
The density of states (\ref{cardyspec}) coincides with Cardy's formula, but now in a different regime of validity, since the derivation of Cardy's formula is only valid for $\Delta \gg c$ at fixed $c$.

The results of HKS can also be stated in terms of the partition function as,
%
%
\be
Z(\beta) \approx \left\{ \begin{matrix}   e^{{\beta c \over 12}}~,\quad \beta >2\pi \cr
e^{{\pi^2 c \over 3\beta}}~,\quad \beta < 2\pi~. \end{matrix}  \right.
\ee
These partition functions coincide with those of thermal AdS or BTZ solutions in the bulk.
We should note that the sparseness condition (\ref{hksineq}) is rather mild, allowing in particular for a stringy growth of states.

\vspace{.2cm}
\noindent
{\bf Assumption 2: Factorization of light correlators}
\vspace{.2cm}

We now make a standard assumption that corresponds to having a weakly coupled low energy field theory in the bulk.    Let $\{ O_i\}$ be a collection of operators whose dimensions are all bounded  in the large $c$ limit.   We assume the asymptotic expansion\footnote{We assume that operators are normalized such that their two-point functions have unit coefficient.}
\be
\langle 0| O_1(x_1) \ldots O_n(x_n)|0\rangle \sim \sum_{k=0}^\infty {1\over c^{k/2}} G_n^{(k)}(x_i)~,\quad c \rt \infty~.
\ee
We also assume that the light operator spectrum can be organized into single-trace and multi-trace operators.  Namely, we have a collection of single-trace operators whose connected $k>2$ point functions amongst each other vanish in the large $c$ limit.  Then we have multi-trace operators whose dimension in the large $c$ limit equals the sum of dimensions of their single-trace constituents, and whose correlators in the large $c$ limit are obtained in terms of their constituents by Wick contractions.

The objects $G_n^{(k)}(x_i)$  computed for single-trace operators are what one computes in the bulk via Witten diagrams in AdS, order by order in the bulk loop expansion.  Following \cite{Heemskerk:2009pn}, we expect that there is a one-to-one correspondence between such CFT correlators that obey crossing constraints and those obtained from theories in AdS.  Note though that since the sparseness condition allows for the bulk theory to be stringy in nature, when we refer to ``Witten diagrams" we are not demanding that the bulk theory necessarily be local with higher derivative interaction terms suppressed at the AdS scale, rather we also admit the case of bulk amplitudes computed from a worldsheet construction with string tension at the AdS scale.  The question of whether a more restrictive sparseness condition, along with modular invariance, implies locality below the AdS scale is discussed in \cite{Belin:2016dcu}.

\vspace{.2cm}
\noindent
{\bf Assumption 3: Growth of light correlators in medium energy states}
\vspace{.2cm}

To state this assumption, we let $\{ O_i\}$ denote a collection of light operators, with $\Delta_i< \Delta_c$, where the cutoff $\Delta_c$ is held fixed as $c\rt \infty$, and we let $O_A$ denote a medium operator, obeying $\Delta_c < \Delta_A < {c\over 12}+\eps$.    Our assumption is that for all sufficiently large $\Delta_c $ and $c$, there exists some number $K$ and positive number $p$ (which are allowed to depend on $n$ and on the positions $x_i$) such that
\be
|\langle A | O_1(x_1) \ldots O_n(x_n) |A\rangle | < K (\Delta_A)^p~.
\ee
Essentially, we are allowing light correlators in medium states to grow with $c$ as long as they do so in a sub-exponential manner.

The bulk intuition behind this assumption is the following.  We expect there to exist bulk solutions with scalar fields taking macroscopic values, but which are not so heavy as to create black holes. Finite gravitational back reaction allows for fields taking values $\phi \sim \sqrt{c}$, so that the matter stress tensor behaves as $T_{\mu\nu} \sim c \sim G_N^{-1}$.  If now we compute the $n$-point function of the CFT operator $O_\phi$ dual to the bulk field $\phi$, we will obtain a result that behaves as $c^{n/2}$.   Our assumption allows for such behavior, but rules out a stronger exponential growth.

\vspace{.2cm}
\noindent
{\bf Assumption 4: Large $c$ expansion of light thermal correlator}
\vspace{.2cm}

Finally we make what is, we believe, a rather mild assumption imposed purely on the light operators, i.e. those with $\Delta < \Delta_c$ as $c\rt \infty$.   Working on the cylinder, we define the light contribution to the thermal correlator as\footnote{We just consider the two-point function here, since that is all we will use, but the generalization is obvious.}
\be
G_2^{(L)}(\phi,t;\beta)  =\frac{e^{\beta c/12}}{Z(\beta)}\sum_L e^{-\beta \Delta_L} \langle L |O(\phi,t)O(0,0)|L\rangle~,\quad \Delta_L, \Delta_O < \Delta_c~.
\ee
For $\beta > 2\pi$, we perform an asymptotic expansion in $1/\sqrt{c}$ (as in Assumption 2), and then take $\Delta_c\rt \infty$ to obtain
\be
G^{(L)}_2(\phi,t;\beta) =  \sum_{k=0}^\infty {1\over c^{k/2} } G_2^{(L,k)}(\phi,t;\beta)~.
\ee
The nontrivial assumption here is the existence of the limit $\Delta_c \rt \infty$ for all the coefficient functions $ G_2^{(L,k)}(\phi,t;\beta)$; in principle it is possible that the functions appearing in the large $c$ expansion grow with $\Delta_c$, as could potentially occur if the matrix elements $\langle L |O(\phi,t)O(0,0)|L\rangle$ are permitted to grow exponentially in $\Delta_c$ for $\Delta_L \sim \Delta_c$.

Given this assumption, the objects $G_2^{(L,k)}(\phi,t;\beta)$ are what one obtains in the bulk computation of the two-point correlator in thermal AdS from Witten diagrams.  In such a computation only light fields propagate in the bulk, with the contribution of virtual heavy states assumed to be exponentially suppressed.

\section{Two-point function analysis}
In this section we show that the Euclidean thermal two-point function of a light operator is fixed by data about the light spectrum. Our result will be
\begin{align}\begin{split}\label{tpr}
\vex{\fO(\phi,t)\fO(0,0)}_{\beta} \approx \begin{cases}
\frac{e^{\beta \frac{c}{12}}}{Z(\beta)} \sum_{A,B\in L} e^{-\beta \D_A}e^{t\D_{AB}}e^{i\phi J_{AB}}|\vex{A|\fO|B}|^2, &\beta>2\pi\\
\frac{e^{\beta' \frac{c}{12}}}{Z(\beta)}\lr{\tfrac{2\pi}{\beta}}^{2\D} \sum_{A,B\in L} e^{-\beta' \D_A}e^{\frac{2\pi \phi}{\beta}\D_{AB}}e^{i\frac{2\pi t}{\beta} J_{AB}}|\vex{A|\fO|B}|^2, &\beta<2\pi
\end{cases}
\end{split}\end{align}
where $\approx$ indicates equality up to corrections that are suppressed exponentially in $\D_c$, $\eps$, or $c$. The summation variables $A$ and $B$ run over a basis of light states with dimensions $\D_A,\D_B$ and spins $J_A,J_B$. $\D_{AB}$ stands for $\D_A-\D_B$ and similarly for $J_{AB}$. All matrix elements lacking a temperature-indicating subscript are to be evaluated on a cylinder (with circumference $2\pi$) and whenever the position of an operator $\fO$ is left implicit it is to be understood as $\phi=0$, $t=0$.

The important point is that the right hand side of equation \eqr{tpr} depends only on the dimensions and OPE coefficients of light operators.

The low temperature case of equation \eqr{tpr} will be established using the assumptions of section \ref{sumps} along with the constraint of modular covariance, which reads
\begin{align}\begin{split}\label{mcv}
\vex{\fO(\phi,t)\fO(0,0)}_{\beta} =  \lr{\frac{2\pi}{\beta}}^{2\D}\vex{\fO(\tfrac{2\pi}{\beta}t,\tfrac{2\pi}{\beta}\phi )\fO(0,0)}_{\beta'},\quad \beta' = \frac{(2\pi)^2}{\beta}~.
\end{split}\end{align}
With the low temperature result in hand, a final application of equation \eqr{mcv} immediately yields the high temperature result.

The starting point of our analysis is the expression for the two-point function at any temperature obtained from cutting the path integral along two time slices that separate the operators $\fO$ and inserting complete sets of states $A,B$,
\begin{align}\begin{split}\label{cc1}
\vex{\fO(\phi,t)\fO(0,0)}_{\beta} = \frac{e^{\beta c/12}}{Z(\beta)}\sum_A\sum_B e^{-\beta \D_A}e^{t\D_{AB}}e^{i\phi J_{AB}}|\vex{A|\fO|B}|^2 ~.
\end{split}\end{align}
The right hand side of equation \eqr{cc1} will sometimes be denoted $G(\phi,t;\beta)$ for brevity.

Given that each state $A,B$ can be light, medium, or heavy, $G(\phi,t;\beta)$ has nine contributions to consider. We will refer to these contributions as $G^{(LL)}, G^{(LM)},...\;G^{(MH)},G^{(HH)}$, where the first superscript refers to $A$ and the second to $B$. We wish to prove the top case of equation \eqr{tpr}, which states that when $\beta>2\pi$  the function $G$ is equal to $G^{(LL)}$ up to exponentially small corrections from the other eight contributions. In subsection \ref{mhlsec} we argue from our assumptions that the seven contributions $G^{(M x)}, G^{(x M)}, G^{(LH)}, G^{(HL)}$ are all small, where $x$ stands for $L$, $M$, or $H$. This leaves $G^{(HH)}$. We then present in subsection \ref{hhsec} an HKS-like argument that modular covariance demands $G^{(HH)}$ to be small.

\subsection{Bounding the medium and off-diagonal contributions}\label{mhlsec}
In this subsection we use the assumptions about the light spectrum laid out in section \ref{sumps} to argue that many of the contributions to equation \eqr{cc1} are small.

\subsubsection{Medium contributions}
The sum of the three quantities $G^{(Mx)}$ is given by
\begin{align}\begin{split}
G^{(ML)}+G^{(MM)}+G^{(MH)} = \frac{e^{\beta c/12}}{Z(\beta)}\sum_{A\in M} e^{-\beta \D_A}\vex{A|\fO(\phi,t)\fO(0,0)|A}
\end{split}\end{align}
where we used the fact that $B$ runs over a complete set of states. Assumptions 1 and 3 above bound the size of this sum as follows:
\begin{align}\begin{split}\label{mba}
\abs{G^{(ML)}+G^{(MM)}+G^{(MH)}}  \leq  \frac{e^{\beta c/12}}{Z(\beta)}\int_{\D_c}^{\infty} d\D\, e^{(2\pi-\beta) \D}K\D^p~.
\end{split}\end{align}
The factor of $e^{2\pi \D}$ comes from the upper bound on the density of states in assumption 1. The upper limit of integration has been set to infinity rather than $c/12+\eps$, a valid step because doing so only makes the inequality weaker. When $\beta>2\pi$ the integral is exponentially small at large $\D_c$ and we conclude that the three contributions\footnote{To proceed from smallness of the left hand side of \eqr{mba} to smallness of the three terms individually, we note that when $\phi=0$ each of the three terms is positive and that taking $\phi$ nonzero can only decrease each term's absolute value.} $G^{(Mx)}$ are exponentially small in $\D_c$.

Exchanging $A$ and $B$ in equation \eqr{cc1} is equivalent to transforming $(\phi,t)$ to $(-\phi,\beta-t)$, so we conclude that $G^{(xM)}$ is also small.

\subsubsection{Heavy-light contributions}
We now argue\footnote{A similar argument appears in \cite{Pappadopulo:2012jk}. Indeed, exponential suppression of $G^{(LH)}$ is a manifestation of their result that the OPE converges exponentially fast.} that the contributions $G^{(LH)}$, $G^{(HL)}$ are suppressed exponentially in $c$ when $\beta>2\pi$. Because of the symmetry in $A,B$ mentioned above, it is sufficient to focus on $G^{(LH)}$. Setting $\phi=0$ for the moment, the contribution in question is
\begin{align}\begin{split}\label{hla}
G^{(LH)}= \frac{e^{\beta c/12}}{Z(\beta)}\sum_{A\in L} \sum_{B\in H}e^{-\beta\D_A}e^{(t-t_0)\D_{AB}}e^{t_0\D_{AB}}\abs{\vex{A|\fO|B}}^2~.
\end{split}\end{align}
A constant $t_0$ in the range $0<t_0<t$ has been introduced to be used in the next step. Obviously the right hand side of equation \eqr{hla} is independent of $t_0$. The factor $e^{(t-t_0)\D_{AB}}$ in the summand above is no larger than $e^{(t-t_0)(\D_c-c/12)}$, and since each term in the sum is positive this implies
\begin{align}\begin{split}\label{hlb}
G^{(LH)}\leq e^{(t-t_0)(\D_c-c/12)}\frac{e^{\beta c/12}}{Z(\beta)}\sum_{A\in L} \sum_{B\in H}e^{-\beta\D_A}e^{t_0\D_{AB}}\abs{\vex{A|\fO|B}}^2~.
\end{split}\end{align}
The inequality only weakens upon extending the sum over $B\in H$ to a sum over all states:
\begin{align}\begin{split}\label{hlc}
G^{(LH)}\leq e^{(t-t_0)(\D_c-c/12)}\frac{e^{\beta c/12}}{Z(\beta)}\sum_{A\in L} e^{-\beta\D_A}\vex{A|\fO(0,t_0)\fO(0,0)|A}~.
\end{split}\end{align}
The first factor on the right hand side is exponentially small. Meanwhile, the rest of that expression is zeroth-order in $c$ by assumption 4. So equation \eqr{hlc} tells us the heavy-light contribution is exponentially small in $c$ relative to $e^{\beta c/12}$. The extension to $\phi\neq 0$ is immediate, because each term in equation \eqr{cc1} only decreases in absolute value upon taking $\phi$ nonzero. We conclude that $G^{(LH)}$, $G^{(HL)}$ are suppressed exponentially in $c$.

\subsection{Heavy-heavy bound from modular covariance}\label{hhsec}
In this subsection, we use modular covariance to show that the contribution to equation \eqr{cc1} from states $A,B$ that are both heavy is suppressed, at $\beta>2\pi$, relative to the full sum. The argument, which parallels that of \cite{Hartman:2014oaa}, is independent of any assumptions about the CFT spectrum, OPE coefficients, or the size of the central charge. All we require is that $\eps$ is large enough for the quantity
\begin{align}\begin{split}\label{dd}
\delta = (\tfrac{\beta}{2\pi})^{2(\D+2)} e^{-|\beta-\beta'|\eps}~
\end{split}\end{align}
to be small. We will find that $G^{(HH)}$ is suppressed relative to $G$ by a factor of $\delta$.

Modular covariance of equation \eqr{cc1} leads to an equivalent expression for the two-point function:
\begin{align}\begin{split}\label{cc2}
\vex{\fO(\phi,t)\fO(0,0)}_{\beta} =(\tfrac{2\pi}{\beta})^{2\D}\frac{e^{\beta' c/12}}{Z(\beta)}\sum_A\sum_B e^{-\beta' \D_A}e^{\frac{\beta'}{2\pi} \phi \D_{AB}}e^{\frac{2\pi}{\beta}itJ_{AB}}|\vex{A|\fO|B}|^2 .
\end{split}\end{align}
We will call the right hand side of \eqr{cc2} $\tG(\phi,t;\beta)$ just as the right hand side of \eqr{cc1} was called $G(\phi,t;\beta)$. The left hand sides of those equations are identical so of course $G(\phi,t;\beta) =  \tG(\phi,t;\beta)$. This is the modular crossing equation. We will find that it puts nontrivial constraints on the matrix elements $\vex{A|\fO|B}$.

We now Fourier expand $G(\phi,t;\beta)$ and $\tG(\phi,t;\beta)$ as functions of $\phi$ and $t$.
\begin{subequations}
\begin{align}
G(\phi,t;\beta) &= \sum_{n=0}^{\infty}\sum_{m=0}^{\infty}G_{nm}\cos(n\phi)\cos (\tfrac{2\pi mt}{\beta})~,\\
\tG(\phi,t;\beta) &= \sum_{n=0}^{\infty}\sum_{m=0}^{\infty}\tG_{nm}\cos(n\phi)\cos (\tfrac{2\pi mt}{\beta})~.
\end{align}
\end{subequations}
Again, the two functions are equal so $G_{nm}=\tG_{nm}$. Only cosine modes appear because the functions are even under the reflections $\phi\to 2\pi-\phi$ and $t\to \beta-t$. Performing the Fourier transforms term by term inside the sums \eqr{cc1}, \eqr{cc2} yields
\begin{subequations}\label{st}
\begin{align}
G_{nm} &= 2\beta \frac{e^{\beta c/12}}{Z(\beta)}\sum_{\substack{A,B\\ |J_{AB}|=n}}
\lr{\frac{ \D_{AB}\lr{e^{-\beta \D_B}-e^{-\beta \D_A}}}{\beta^2 \D_{AB}^2+(2\pi m)^2}}
|\vex{A|\fO|B}|^2\label{s1}\\
\tG_{nm} &= 2\beta' \lr{\frac{2\pi}{\beta}}^{2\D}\frac{e^{\beta' c/12}}{Z(\beta)}\sum_{\substack{A,B\\ |J_{AB}|=m}}
\lr{\frac{ \D_{AB}\lr{e^{-\beta' \D_B}-e^{-\beta' \D_A}}}{{\beta'}^2 \D_{AB}^2+(2\pi n)^2}}
|\vex{A|\fO|B}|^2~.\label{t1}
\end{align}
\end{subequations}
When $\D_A=\D_B$ the summands of \eqr{s1}, \eqr{t1} are to be defined via their limits as $\D_A\to \D_B$, which are finite and nonnegative.

Note that the contribution from any pair of states $A,B$ to either sum \eqr{s1}, \eqr{t1} is nonnegative. This fact is central to the argument below, which parallels the original one applied by HKS to the partition function. We begin by separating out the heavy-heavy contribution to each sum \eqr{cc1},\eqr{cc2}:
\begin{subequations}
\begin{align}
G(\phi,t;\beta)&=G^{(L)}(\phi,t;\beta)+G^{(HH)}(\phi,t;\beta)\\
\tG(\phi,t;\beta)&=\tG^{(L)}(\phi,t;\beta)+\tG^{(HH)}(\phi,t;\beta)~.
\end{align}
\end{subequations}
We define $G^{(HH)}$ to be the contribution to \eqr{cc1} from states $A,B$ that are both heavy, as above. Meanwhile $G^{(L)}$ is the contribution from all other pairs of states\footnote{Note that $G^{(L)}$ is distinct from $G^{(LL)}$ (although the conclusion of subsection \ref{mhlsec} is that these functions' difference is small).}. $\tG^{(HH)}$, $\tG^{(L)}$ are defined in the same way.

The Fourier modes of $G^{(HH)},\tG^{(HH)}$ are found by restricting sums \eqr{st} to heavy states:
\begin{subequations}
\begin{align}
G_{nm}^{(HH)} &= 4\beta \sum_{A\in H}\sum_{\substack{B,\, \D_B\geq \D_A\\ |J_{AB}|=n}}(\tfrac{1}2)^{\delta_{AB}}
\frac{e^{-\beta E_A}}{Z(\beta)}\lr{\frac{ -\D_{AB}\lr{1-e^{\beta \D_{AB}}}}{\beta^2 \D_{AB}^2+(2\pi m)^2}}
|\vex{A|\fO|B}|^2\label{t2}\\
\tG_{nm}^{(HH)} &= 4\beta' (\tfrac{2\pi}{\beta})^{2\D}\sum_{A\in H}\sum_{\substack{B,\, \D_B\geq \D_A\\ |J_{AB}|=n}} (\tfrac{1}2)^{\delta_{AB}}\frac{e^{-\beta' E_A}}{Z(\beta)}
\lr{\frac{ -\D_{AB}\lr{1-e^{\beta' \D_{AB}}}}{{\beta'}^2 \D_{AB}^2+(2\pi n)^2}}
|\vex{A|\fO|B}|^2~.\label{x2}
\end{align}
\end{subequations}
Above we used the symmetry in $A,B$ of equations \eqr{st} to arrange for $\D_A$ to be less than or equal to $\D_B$ inside the sum. For proper counting we introduced a factor of $2(1/2)^{\delta_{AB}}$ which is 1 if $A=B$ and $2$ otherwise. (There is no loss of generality in assuming $B$ runs over the same basis as $A$.)

It's important to keep in mind for what follows that $\beta> 2\pi$. Following HKS we note that we can bound the Boltzmann factor in equation \eqr{t2} for $G^{(HH)}_{nm}$ in the following way
\begin{align}\begin{split}
e^{-\beta E_A} = e^{-(\beta-\beta') E_A}e^{-\beta' E_A} \leq e^{-(\beta-\beta') \eps}e^{-\beta' E_A}~.
\end{split}\end{align}
This is the step in which it is important that the lightest heavy state has energy $\eps$ larger than zero. We then use the fact that every term in the sum $G^{(HH)}_{nm}$ is positive to write
\begin{align}\begin{split}\label{tr1}
G_{nm}^{(HH)} \leq 4\beta e^{-(\beta-\beta')\eps}\sum_{\substack{A\\ \text{heavy}}}\sum_{\substack{B,\, \D_B\geq \D_A\\ |J_{AB}|=n}} (\tfrac{1}2)^{\delta_{AB}}\frac{e^{-\beta' E_A}}{Z(\beta)}
\lr{\frac{ -\D_{AB}\lr{1-e^{\beta \D_{AB}}}}{{\beta}^2 \D_{AB}^2+(2\pi n)^2}}
|\vex{A|\fO|B}|^2~.
\end{split}\end{align}
The quantity in parentheses in \eqr{tr1} is bounded by
\begin{align}\begin{split}\label{gb}
\lr{\frac{ -\D_{AB}\lr{1-e^{\beta \D_{AB}}}}{{\beta}^2 \D_{AB}^2+(2\pi n)^2}}
\leq \lr{\frac{\beta}{\beta'}} \lr{\frac{ -\D_{AB}\lr{1-e^{\beta' \D_{AB}}}}{{\beta'}^2 (\D_{AB}^2+(2\pi n)^2}}.
\end{split}\end{align}
To check \eqr{gb}, first note that the denominator on the left is $\geq$ the one on the right. And second note that the fraction $(1-e^{\beta \D_{AB}})/(1-e^{\beta' \D_{AB}})$ is an increasing function of $\D_{AB}<0$ and so is bounded from above by its limit as that difference goes to zero, which is $\beta/\beta'$. From \eqr{gb} it follows that
\begin{align}\begin{split}\label{tr2}
G_{nm}^{(HH)} \leq  \tfrac{4\beta^2}{\beta'} e^{-(\beta-\beta')\eps} \sum_{A\in H}\sum_{\substack{B,\, \D_B\geq \D_A\\ |J_{AB}|=n}} (\tfrac{1}2)^{\delta_{AB}}\frac{e^{-\beta' E_A}}{Z(\beta)}
\lr{\frac{-\D_{AB}\lr{1-e^{\beta' \D_{AB}}}}{{\beta'}^2 \D_{AB}^2+(2\pi n)^2}}
|\vex{A|\fO|B}|^2.
\end{split}\end{align}
We recognize the right hand side as proportional to $\tG_{mn}^{(H)}$ and conclude that
\begin{align}\begin{split}\label{TSineq}
G_{nm}^{(HH)}\leq \delta \tG_{mn}^{(HH)}
\end{split}\end{align}
with $\delta$ being the constant defined in equation \eqr{dd}.

Equation \eqr{TSineq} is analogous to the starting point, equation \eqr{HKS2.6}, of HKS's analysis for the partition function: $G^{(HH)}$, $\tG^{(HH)}$ and $\delta$ play the roles of $Z_H$, $Z_H'$ and $e^{(\beta'-\beta)\eps}$, respectively. From that starting point a clever modular invariance argument (reviewed in section \ref{HKSrev}) showed that the heavy contribution to the low temperature partition function is suppressed relative to the light contribution.  In appendix \ref{vin} we apply the same argument to the two-point function's Fourier modes. The result is
\begin{align}\begin{split}\label{ter6}
G_{nm}^{(HH)} \leq \delta \frac{G_{mn}^{(L)}+\delta G_{nm}^{(L)}}{1-\delta^2}~.
\end{split}\end{align}

The right hand side of this result is suppressed by a factor of $\delta$ relative to $G_{mn}^{(L)}$ and $G_{nm}^{(L)}$, which are of order unity by the results of section \ref{mhlsec}. We conclude that $G_{nm}^{(HH)}$ is suppressed by $\delta$ relative to unity. That this holds for every Fourier mode implies the function $G^{(HH)}(\phi,t;\beta)$ is itself suppressed by a factor of $\delta$. That is, it is exponentially small in the parameter $\eps$.

\subsection{Conclusion of this section}

From the results of subsections \ref{mhlsec} and \ref{hhsec},  equation \eqr{tpr} is established. At $\beta>2\pi$, the right hand side of that equation is what one computes using Witten diagrams in thermal AdS. To be precise, if one expands the right hand side in powers of $\frac{1}{\sqrt{c}}$ and then takes the limit $\D_c\to \infty$ order-by-order one recovers the Witten diagram expansion in powers of $\frac{1}{\sqrt{c}}$. At $\beta<2\pi$ the same conclusion holds with thermal AdS replaced by a BTZ black hole of the appropriate temperature. This follows from the low temperature statement by bulk modular invariance: The Euclidean BTZ black hole of inverse temperature $\beta$ is isometric to thermal AdS at inverse temperature $\beta'$, and the isometry involves exchanging the angular and time coordinates as in equation \eqr{mcv}.

\section{One-point function}

With our two-point function result in hand, we now turn to the one-point function.  We first derive an upper bound on the one-point function.   We start with the decomposition of the two-point function,
\bea
\langle O(0,\beta/2)O(0,0)\rangle_\beta = {1\over Z(\beta)} \sum_{A,B} e^{-{\beta \over 2}(E_A+E_B)} |\langle A|O|B\rangle|^2~.
\eea
The right hand side is bounded below by the contribution from $A=B$, hence
\bea
\langle O(0,\beta/2)O(0,0)\rangle_\beta \geq  {1 \over Z(\beta)} \sum_{A} e^{-{\beta}E_A} |\langle A|O|A\rangle|^2~.
\eea
Now, using the elementary fact that $x^2 \geq 2x-1$ for real $x$, we obtain
\bea
\langle O(0,\beta/2)O(0,0)\rangle_\beta \geq   {2 \over Z(\beta)} \sum_{A} e^{-{\beta}E_A} |\langle A|O|A\rangle|-1~.
 \label{oneeq}
\eea
(\ref{oneeq}) implies a bound on the contribution to the one-point function from any collection of states,
\bea
 {1 \over Z(\beta)} \sum_{A \in \psi} e^{-{\beta}E_A} |\langle A|O|A\rangle| \leq  { 1 +  \langle O(0,\beta/2)O(0,0)\rangle_\beta \over 2}~,
 \label{onebound}
\eea
where $\psi$ denotes any subspace of the full Hilbert space.
Since $|\langle O \rangle_\beta| \leq {1\over Z(\beta)} \sum_A e^{-\beta E_A} |\langle A|O|A\rangle|$ we also deduce a bound on the full one-point function
\bea
|\langle O \rangle_\beta|  \leq { 1 + \langle O(0,\beta/2)O(0,0)\rangle_\beta \over 2}~.
\label{aaa}
\eea
These are exact bounds, valid in any theory for any temperature.   For our purposes, the main fact we will use is that since the two-point function is finite in the large $c$ limit, the same is true of the left hand side of (\ref{onebound}).

We now show that for $\beta >2\pi$  the contributions to the one-point function from medium and heavy states are exponentially suppressed compared to the light state contribution.   For the medium state contribution we proceed as in (\ref{mba}),
\bea
 \left| {1\over Z(\beta)} \sum_{A\in M} e^{-\beta E_A} \langle A|O|A\rangle \right|  &\leq &  {1\over Z(\beta)} \sum_{A\in M}  e^{-\beta E_A} |\langle A|O|A\rangle|
 \cr & \leq & {e^{\beta c\over 12} \over Z(\beta)} \int_{\Delta_c}^\infty \! d\Delta e^{(2\pi - \beta) \Delta} K \Delta^p~.
\eea
For $\beta> 2\pi$ the last expression is exponentially small in $\Delta_c$.   For the heavy contribution we note that
\bea
 \left| {1\over Z(\beta)} \sum_{A\in H} e^{-\beta E_A} \langle A|O|A\rangle \right|  &\leq &  {1\over Z(\beta)} \sum_{A\in H}  e^{-\beta E_A} |\langle A|O|A\rangle|~
 \eea
together with (\ref{onebound}) implies that the heavy contribution has a finite large $c$ limit.   In particular, this holds for any $\beta$ sufficiently greater than $2\pi$ such that $\delta \ll 1$.  But then for any larger $\beta$ the right hand side is exponentially small in $c$, since $Z(\beta)^{-1} e^{-\beta E_A}$ is exponentially small for all heavy states.

Since the medium and heavy state contributions are exponentially small, we conclude that
\bea
\langle O \rangle_\beta \approx {1\over Z(\beta)}  \sum_{A\in L} e^{-\beta E_A} \langle A|O|A\rangle~,  \quad \beta > 2\pi
\eea
as desired.  A modular transformation then gives the high temperature result,
\bea
\langle O \rangle_\beta \approx  \left(2\pi \over \beta\right)^\Delta {1\over Z(\beta)}  \sum_{A\in L} e^{-{4\pi^2 \over \beta} E_A} \langle A|O|A\rangle~,  \quad \beta < 2\pi
\eea
Since the three-point coefficients $\langle A|O|A\rangle $  of single-trace operators fall off at least as fast as $1/\sqrt{c}$ according to Assumption 2, it follows that the generic one-point function of a single-trace operator is $O(1/\sqrt{c})$.  Double-trace operators can have $O(1)$ expectation values, in accord with (\ref{aaa}).

In \cite{Kraus:2016nwo} it was pointed out that modular invariance implies that the one-point function in the high temperature limit behaves as
\bea
\langle O \rangle_\beta \approx  \left(2\pi \over \beta\right)^\Delta {1\over Z(\beta)}  e^{-{4\pi^2 \over \beta} E_\chi} \langle \chi |O|\chi \rangle~,  \quad \beta \rt 0~,
\eea
where $|\chi\rangle$ denotes the lightest state such that $\langle \chi |O|\chi \rangle \neq 0$.  This results holds for all $c$. It was further noted that this asymptotic formula for the one-point function has a simple bulk interpretation in terms of a $\chi$ particle winding around the BTZ horizon.    What we have shown here is that in a large $c$ theory satisfying our assumptions, the analogous result holds, except now we should sum over all light states winding around the horizon, including multiparticle states.  This then yields the one-point function for all $\beta <2\pi$.

\section{Discussion}

We conclude with a few comments.

Our main CFT result is that, under our assumptions, one and two point correlators of light fields at any temperature are determined entirely by light spectrum data.
Translated into bulk gravity language, the statement is that thermal AdS and the BTZ black hole emerge as the universal backgrounds for the computation of thermal one and two-point functions of light operators, and only the propagation of light fields on these backgrounds need be considered.  One obvious extension is to generalize to n-point functions of light operators.  We anticipate no fundamental obstacles here, although the analysis will become more complicated.

Our result for the thermal two-point function assumes that the time separation is purely Euclidean and is held fixed as $c  \rt \infty$.  Indeed, we expect our results to breakdown if we instead allow for Lorentzian time separations that can grow with $c$, since in this case we would otherwise violate bounds on the size of such correlators: perturbative Witten diagrams yield a result that decays exponentially to zero at late times, whereas unitarity places a lower bound on the long time average \cite{Maldacena:2001kr}.   It would be interesting to explore in detail the regime of validity of our results once we relax the conditions on the time arguments.

It may be instructive to compute correlators in symmetric product orbifold theories as an explicit realization of our assumptions.  Such theories are known \cite{ Hartman:2014oaa} to saturate the density of states allowed by the HKS analysis, and their correlators admit a $1/\sqrt{c}$ expansion.  Of course, such theories, being free, are far from having a bulk description in terms of Einstein gravity (for example, they are not chaotic \cite{Belin:2017jli}), but this is the price paid for calculability on the CFT side.  Some related computations were carried out in \cite{Balasubramanian:2005qu}.

It would be very interesting if analogous statements to what we have shown here could be established in higher dimensions.  On general grounds we expect a similar result to hold, but it is clear that new issues arise.  Namely, modular invariance acts in a more complicated way in higher dimensions, relating CFTs on distinct spaces to one another rather than just changing the temperature \cite{Shaghoulian:2015kta,Shaghoulian:2015lcn,Belin:2016yll,Shaghoulian:2016gol}. The bulk analog of this statement is that the black hole solution ---  AdS-Schwarzschild --- is no longer locally AdS as for BTZ, but depends on the details of the bulk theory, such as the presence of higher derivative terms and so on. The story will thus necessarily be more intricate.

\section*{Acknowledgments}

We thank Tom Hartman, Alex Maloney, and Eric Perlmutter for useful discussions.   P.K. is supported in part by NSF grant PHY-1313986.

\appendix

\section{Details of modular crossing analysis}\label{vin}
In subsection \ref{hhsec} we showed that the heavy contributions to the Fourier modes of the two-point function satisfy an inequality \eqr{TSineq}, which we repeat here for convenience:
\begin{equation}\label{pendx1}
G_{nm}^{(HH)} \leq \delta \tG_{mn}^{(HH)}~.
\end{equation}
The indices $n,m$ are transposed between the left and right hand sides. To treat this complication it is convenient to combine the above relation and its image under $n\leftrightarrow m$ into a single inequality:
\begin{equation}\label{vecinv}
\lr{\begin{array}{c} G_{nm}^{(HH)}\\
G_{mn}^{(HH)}
\end{array}}\leq
\lr{\begin{array}{cc}0 &\delta \\ \delta &0
\end{array}}
 \lr{\begin{array}{c} \tG_{nm}^{(HH)}\\
\tG_{mn}^{(HH)}
\end{array}}.
\end{equation}
We use a vector inequality such as \eqr{vecinv} to indicate that the comparison holds separately for each component.

Our argument so far has not invoked modular invariance. That is, we have not yet used the fact that $G_{nm}^{(HH)}+G_{nm}^{(L)} = \tG_{nm}^{(HH)}+\tG_{nm}^{(L)}$. We use it now. Since $\tG_{nm}^{(L)}$ is nonnegative, modular invariance implies  $G_{nm}^{(L)}\geq \tG_{nm}^{(HH)}-G_{nm}^{(HH)}$, which in light of equation \eqr{vecinv} implies

\begin{equation}\label{ter1}
\lr{\begin{array}{c} G_{nm}^{(L)}\\
G_{mn}^{(L)}
\end{array}}\geq
\lr{\begin{array}{cc}1 &-\delta\\ -\delta & 1
\end{array}}
 \lr{\begin{array}{c} \tG_{nm}^{(HH)}\\
\tG_{mn}^{(HH)}
\end{array}}.
\end{equation}
One can multiply both sides of a strict vector inequality by a matrix as long as every element of the matrix is nonnegative. In particular as long as $|\delta| <1$ we can multiplying both sides of \eqr{ter1} by the inverse of the matrix on the right hand side. The result is
\begin{equation}\label{ter2}
 \lr{\begin{array}{c} \tG_{nm}^{(HH)}\\
\tG_{mn}^{(HH)}
\end{array}}
\leq
\frac{1}{1-\delta^2}\lr{\begin{array}{cc}1 &\delta \\ \delta & 1
\end{array}}\lr{\begin{array}{c} G_{nm}^{(L)}\\
G_{mn}^{(L)}
\end{array}}.
\end{equation}
Inequality \eqr{ter2} is the analog of \eqr{HKS2.8} in the HKS argument. We now substitute \eqr{ter2} into the right hand side of the original inequality \eqr{vecinv}, a valid step because the matrix there has nonnegative elements, to get
\begin{equation}\label{ter5}
\lr{\begin{array}{c} G_{nm}^{(HH)}\\
G_{mn}^{(HH)}
\end{array}}\leq
\frac{\delta}{1-\delta^2}
\lr{\begin{array}{cc}\delta & 1\\ 1 & \delta
\end{array}}\lr{\begin{array}{c} G_{nm}^{(L)}\\
G_{mn}^{(L)}
\end{array}}.
\end{equation}
The above relation is analogous to \eqr{HKS2.9} in the HKS argument. The upper component is
\begin{equation}\label{pendxn}
G_{nm}^{(HH)} \leq \delta \frac{G_{mn}^{(L)}+\delta G_{nm}^{(L)}}{1-\delta^2}~,
\end{equation}
which is equation \eqr{ter6}.

\bibliographystyle{ssg}
\bibliography{biblio}

\end{document}